\documentclass[showpacs,aps,amssymb,pra,amsmath]{revtex4}
\usepackage{float}
\usepackage{epsfig}
\usepackage{hyperref}
\usepackage{bm}
\usepackage{amsmath,graphicx}
\usepackage{subfig}
\usepackage{setspace}
\usepackage{color}

\newcommand{\beq}{\begin{equation}}
\newcommand{\eeq}{\end{equation}}
\newcommand{\bea}{\begin{eqnarray}}
\newcommand{\eea}{\end{eqnarray}}

\def\bq{{\bf q}}

\def\br{{\bf r}}

\def\bp{{\bf p}}
\def\bs{{\bf s}}
\def\bR{{\bf R}}

\begin{document}
\title{A manifestly Hermitian semiclassical expansion for the one-particle  density matrix of a two-dimensional Fermi gas}
\author{K. Bencheikh$^{(1)}$, B. P. van Zyl$^{(2)}$, and K. Berkane$^{(1)}$}
\affiliation{$^{(1)}$Laboratoire de Physique
Quantique et Syst\`{e}mes Dynamiques. D\'{e}partement de Physique.Universit\'{e} Ferhat Abbas S\'{e}tif-1, and Setif 19000, Algeria,}
\affiliation{$^{(2)}$Department of Physics, St. Francis Xavier University, Antigonish, Nova Scotia, Canada B2G 2W5}
\date{\today}

\begin{abstract}
The semiclassical $\hbar$-expansion of the one-particle density matrix for a two-dimensional Fermi gas is calculated within the Wigner transform method of 
Grammaticos and Voros, originally developed  in the context of nuclear physics.   
The method of Grammaticos and Voros has the virture of preserving both the Hermiticity and idempotency of the density matrix to all orders in the 
$\hbar$-expansion.
As a topical application, we use our semiclassical expansion to go beyond the local-density
approximation for the construction of the total dipole-dipole interaction energy functional of a 
two-dimensional, spin-polarized dipolar Fermi gas.  We find a {\em finite},  second-order gradient correction to the Hartree-Fock energy, which takes the form
$\varepsilon (\nabla \rho)^2/\sqrt{\rho}$, with $\varepsilon$ being small ($|\varepsilon| \ll1$) and negative.  We test the quality of the corrected energy by
comparing it with the 
exact results available for harmonic confinement.  Even for small particle numbers, the gradient correction to the dipole-dipole energy
provides a significant improvement over the local-density approximation.
\end{abstract}
\maketitle

\section{Introduction}\label{intro}

Density-functional theory (DFT)~\cite{DFT} is by far the most common and powerful numerical approach for the solution of the quantum many-body problem of $N$ interacting fermions, and constitutes the cornerstone for research
in diverse fields such as condensed matter and nuclear
physics, quantum chemistry, and materials science.  
Specifically, the  Hohenberg-Kohn-Sham (HKS) DFT~\cite{DFT} states that the ground state properties of an $N$-body interacting Fermi system may be mapped to a noninteracting 
system of independent fermions
moving in an effective one-body potential, $v_{\rm eff}(\br)$, sometimes referred to as the Kohn-Sham potential, $v_{\rm KS}(\br)$.   
The HKS total energy functional
is then given by (hereby, we focus to strictly two-dimensional (2D) systems)
\beq\label{functional}
E[\rho] = T_0[\rho] + E_{\rm int}[\rho] + \int d^2r~v_{\rm ext}(\br)\rho(\br)~.
\eeq
In Eq.~\eqref{functional}, $T_0[\rho]$ is the kinetic energy (KE)  of a noninteracting Fermi gas, $E_{\rm int}[\rho]$ accounts for both classical and quantum interactions, and the last term 
is the energy functional associated with the external potential,
$v_{\rm ext}(\br)$.   

The noninteracing KE functional is treated {\em exactly} in the HKS formalism, and by definition is given by (in this paper, we deal with fermions with spin degeneracy $g=1,2$)
\beq\label{T_0}
T_0[\rho]= g \sum_{i=1}^{N/g}\int d^2r~\phi_i^*(\br)\left(-\frac{\hbar^2}{2m}\nabla^2\right)\phi_i(\br)~,
\eeq
where the summation is over fully occupied orbitals, $\{\phi_i(\br)\}$.
The variational minimization of $E[\rho]$ with respect to the density then leads to the following set of  single-particle Schr\"odinger-like equation for the orbitals, $\{\phi_i(\br)\}$,
\beq\label{orbitals}
-\frac{\hbar^2}{2m} \nabla^2 \phi_i(\br) + v_{\rm eff}(\br) \phi_i(\br) = \varepsilon_i \phi_i(\br)~,~~~~(i=1,...,N)~,
\eeq
where the effective potential mentioned above is given by
\beq\label{veff}
v_{\rm eff}(\br) = \frac{\delta E_{\rm int}[\rho]}{\delta \rho} + v_{\rm ext}(\br)~.
\eeq
Therefore, in the HKS scheme, one must self-consistently solve for $N$ Schr\"odinger-like equations, which at self-consistency leads to the 
spatial density 
\beq\label{Shamden}
\rho(\br) = g \sum_{i=1}^{N/g} \phi_i^*(\br)\phi_i(\br)~,
\eeq
with the normalization,
\beq\label{norm}
\int d^2r~\rho(\br) = N~,
\eeq
also determining the Fermi energy, $E_F$.

The  interaction energy functional,  $E_{\rm int}[\rho]$, is generally not known, so some approximation must be made for $E_{\rm int}[\rho]$ in order to completely specify the HKS
functional, Eq.~\eqref{functional}.
Despite its conceptual appeal, any practical implementation of the HKS theory, as defined above, must be weighed against the numerically expensive self-consistent solution to 
$N$ single-particle Schr\"odinger-like equations, Eq.~\eqref{orbitals}.

Ideally, one would like to keep in the original spirit of DFT, in which there is no need for the calculation of  single-particle orbitals of any kind.  In principle, this so-called orbital-free DFT  
can be accomplished within the HKS scheme if
one could construct an explicit density functional for the exact, noninteracting KE, $T_0[\rho]$, for an arbitrary inhomogeneous system.  To date, such a functional has not been found, implying
that if one would like to avoid calculating single-particle orbitals, an additional layer of approximation must be made; that is, in an orbital-free DFT, two functionals, $T_0[\rho]$ and $E_{\rm int}[\rho]$,
must be approximated.  Nevertheless, if the approximation to $T_0[\rho]$ is accurate, the computational cost savings for investigating $N\gg1$ systems is extremely compelling, and is often the
only practical numerical option (e.g., in materials science, where $N \sim {\cal O}(10^{23})$).

Toward the goal of obtaining an expression for the KE functional, $T_0[\rho]$,  explicitly in terms of the spatial density, one may introduce the one-body density matrix (ODM),
which is formally defined in terms of the normalized many-body wave function~\cite{DFT}, $\psi$, viz.,
\beq\label{ODM}
\rho_1(\br;\br') =   N \int d^2r_2...d^2r_N~\psi^*(\br,\br_2,...,\br_N)\psi(\br',\br_2,...,\br_N)~.
\eeq
Note that by definition, the spatial density is given by the diagonal element of the ODM, viz., $\rho(\br) = \rho_1(\br;\br)$, along with the fact that the ODM is Hermitian,
$\rho_1(\br;\br') = [\rho_1(\br';\br)]^*$ .  It will prove advantageous later to introduce  the center-of-mass and relative coordinates, $\bR=(\br+\br')/2$ and
$\bs = \br-\br'$, respectively.  We also define
\beq
 \bar{\rho}_1(\bR;\bs) \equiv \rho_1(\bR + \bs/2;\bR - \bs/2)~,
\eeq
in which case Hermiticity takes the form $\bar{\rho}_1(\bR;\bs)=[\bar{\rho}_1(\bR;-\bs)]^*$
The noninteracting  KE functional is then obtained from
\beq\label{SCKE}
T_0[\rho] = -\frac{\hbar^2}{2m}\int d^2R~[\nabla^2_\bs \bar{\rho}_1(\bR;\bs)]_{\bs = 0}.
\eeq
Unsurprisingly, there is no known explicit expression for the
ODM for an arbitrary inhomogeneous system, implying that approximations to the ODM are unavoidable.
It is evident that the quality of any approximation to the noninteracting KE functional  is inextricably connected with the approximation applied to the ODM.

The crudest expression for  $\bar{\rho}_1(\bR;\bs)$ is the so-called local-density approximation (LDA) in which the form of the ODM of a spatially uniform 2D system is assumed to be locally
valid for an inhomogeneous system~\cite{brack_bhaduri,vanzyl2003},viz.,
\beq\label{2DODM}
\bar{\rho}_1(\bR;\bs) = g \frac{k_F(\bR)^2}{2\pi}\frac{J_1[k_F(\bR)|\bs|]}{k_F(\bR)|\bs|}~,
\eeq
where $k_F(\bR) = \sqrt{4 \pi \rho(\bR)/g}$ is the local Fermi wave vector and $J_n(x)$ is a cylindrical Bessel function of the $n$-th order~\cite{GR}.   It is immediately seen that
Eq.~\eqref{2DODM} is Hermitian.  Note that in, e.g., the Kirzhnits commutator formalism~\cite{DFT}, the fact that one begins with a representation in $\br$ and $\br'$, means that 
the resulting Kirzhnits  LDA for the ODM is not 
Hermitian.    Specifically, in the Kirzhnits approach, one obtains
\beq\label{LDAK}
{\rho}_1(\br;\br') = g \frac{k_F(\br)^2}{2\pi}\frac{J_1[k_F(\br)|\bs|]}{k_F(\br)|\bs|}~,
\eeq
which clearly does not possess the correct $\br$ and $\br'$ symmetry.
However, upon the appropriate symmetrization, one is lead to an expression identical to 
 Eq.~\eqref{2DODM}.
 
Inserting Eq.~\eqref{2DODM} into
Eq.~\eqref{SCKE} leads to the LDA for the noninteracting kinetic energy functional, viz.,
\beq\label{KELDA}
T_0[\rho] =\frac{\pi}{g} \frac{\hbar^2}{m}\int d^2R~\rho(\bR)^2~,
\eeq
which is obviously an explicit functional of the spatial density.  Presumably, going beyond the LDA for the ODM will lead to a more accurate noninteracting KE functional.  We will come
back to this point later in the paper.

The interaction energy functional, $E_{\rm int}[\rho]$, may also be determined solely in terms of the ODM if one adopts the common Hartree-Fock approximation (HFA), which does not take
into account correlations, viz.,
\beq\label{E_intHFA}
E_{\rm int}[\rho] = \int\int d^2rd^2r'~  \frac{1}{2}\left(\rho_1(\br,\br)\rho_1(\br',\br') - \frac{1}{g}\rho_1(\br,\br')\rho_1(\br',\br)\right) v_{\rm int}(\br,\br')~,
\eeq
where $v_{\rm int}(\br,\br')$ is the two-body interaction potential (e.g., a $1/|\br-\br'|$ Coulomb potential). The first term in Eq.~\eqref{E_intHFA} corresponds to the classical
Hartree energy, while the second term represents the quantum-mechanical exchange energy.
We then see that the ODM is not only fundamental for obtaining  the noninteracting KE functional, but also (at least within the HFA) the interaction energy functional.
Unfortunately, to our knowledge, the only inhomogeneous systems for which an exact analytical expresion for the ODM is available are the so-called Bardeen model~\cite{bardeen}, the three-dimensional  harmonic oscillator (HO) with smeared occupancy~\cite{bhaduri}, 
and the multi-dimensional HO~\cite{vanzyl2003,march}.   The point to be taken here is that if one wishes to find explicit functionals for $T_0[\rho]$, and $E_{\rm int}[\rho]$,   approximations to the ODM
must be employed, since $\rho_1(\br;\br')$ is not known exactly for an arbitrary inhomogenous system. 

To this end, in Sec.~\ref{SCexp}, we briefly review the 
Grammaticos-Voros (GV) semiclassical (SC) $\hbar$-expansion for the ODM~\cite{GM}, and subsequently apply it to develop a Hermitian, idempotent, $\hbar$-expansion for the 2D density matrix
of an arbitrary inhomogeneous 2D Fermi gas.  While an analogous calculation of this kind (i.e., using the GV approach) 
has been recently performed by Bencheikh and R\"as\"anen~\cite{BR} in three-dimensions (3D), 
we feel that a presentation of the 2D
analysis is a worthwhile endeavor.  First, from a pedagogical point of view, the calculations involved in obtaining the SC 2D ODM are somewhat unwieldy, so providing the details of such
calculations will be useful to other researchers wishing to apply, or extend our results.  In addition, providing an explicit expression for the SC 2D ODM is of academic interest, since its presentation
utilizing the GV approach is currently not available in the literature.  Finally, the fundamental role of the ODM in applications of DFT to degenerate Fermi gases, suggests that our work will also be of practical importance in diverse areas of research (e.g., instabilities in 2D dipolar Fermi gases (dFG), Wigner crystallization in 2D electronic and dFG,
physics of metal clusters, etc). 

 Following this development, we will apply our results to construct a beyond the LDA expression
for the total HF dipole-dipole interaction energy for a 2D spin-polarized dipolar Fermi gas.  
Our paper closes with a summary and suggestions for 
future work.

\section{Semiclassical $h$-bar expansion of the density matrix}\label{SCexp}
We begin by considering a 2D system of noninteracting fermions under the influence of some
one-body potential, 
\beq\label{OBH}
\hat{H} = -\frac{\hbar^2}{2m} \nabla^2 + V(\br)
\eeq
along with the associated one-particle density operator
\beq\label{ODO}
\hat{\rho} = \Theta(E_F - \hat{H})~,
\eeq
where $E_F$ is the Fermi energy and $\Theta$ is the Heaviside function. 

We develop our semiclassical expansion by working with the Wigner transform of $\rho_1(\br;\br')$,
as originally developed by GV~\cite{GM} and recently applied to the 3D ODM by Bencheikh and 
R\"as\"anen~\cite{BR}
viz.,
\beq\label{WTODM}
\bar{\rho}^{\rm sc}_1(\bR;\bs) =\frac{1}{(2\pi\hbar)^2}\int d^2p~\rho^{\rm sc}_{\rm w}(\bR,\bp) e^{i\bp\cdot\bs/\hbar}~,
\eeq
where $\bp$ is the momentum conjugate to $\bs$, and $\rho_{\rm w}(\bR,\bp)$ denotes the Wigner
transform of $\rho_1(\br;\br')$.  Note that in the GV approach, one immediately works in the $\bR$ and $\bs$ representation, which
then leads to a transparent expression from which one can deduce the Hermiticity of the ODM.

At the heart of the method is to expand 
$\rho_{\rm w}(\bR,\bp)= [\theta(E_F - \hat{H}) ]_{\rm w}$  around the identity operator times
the classical Hamiltonian $H_{\rm cl}$, 
\beq\label{Hcl}
H_{\rm cl} = \frac{\bp^2}{2m}+V(\bR)~.
\eeq
To second order in $\hbar$, one finds the dimensionally independent expression for $\rho^{\rm sc}_{\rm w}(\bR,\bp)$, which reads~\cite{GM,BR}
\beq\label{WT2}
\rho^{\rm sc}_{\rm w}(\bR,\bp) = \Theta(E_F-H_{\rm cl}) - \frac{1}{2}\varphi_2\delta'(H_{\rm cl}-E_F)
-\frac{1}{6}\varphi_3\delta''(H_{\rm cl}-E_F) + {\cal O}(\hbar^4)~,
\eeq
\beq\label{phi2}
\varphi_2 = -\frac{\hbar^2}{4m}\nabla^2_\bR V + {\cal O}(\hbar^4)~,
\eeq
and
\beq\label{phi_3}
\varphi_3 = -\frac{\hbar^2}{4m}\left[ (\nabla_\bR V)^2 + \frac{1}{m}(\bp\cdot\nabla_\bR)^2V\right] +
{\cal O}(\hbar^4)~,
\eeq
where primes refer to derivatives with respect to $H_{\rm cl}$~\cite{note1}.  The semiclassical $\hbar$-expansion
of the ODM is then obtained by inserting Eq.~\eqref{WT2}, along with the expressions
for $\varphi_2$ and $\varphi_3$ into Eq.~\eqref{WTODM}, viz.,
\beq\label{SCODM}
\bar{\rho}^{\rm sc}_1(\bR;\bs) = (A + B + C + D)~,
\eeq
where we recall that $\bar{\rho}_1(\bR;\bs) \equiv \rho_1(\bR+\frac{1}{2}\bs;\bR - \frac{1}{2}\bs)$,
and
\bea\label{ABCD}
A &=& \frac{g}{(2\pi\hbar)^2}\int d^2p~ e^{i\bp\cdot\bs/\hbar}\Theta(E_F-H_{\rm cl})~,\\
B &=& \frac{g}{(2\pi\hbar)^2}\frac{\hbar^2}{8m}\nabla_\bR^2V\int d^2p~ e^{i\bp\cdot\bs/\hbar}\delta'(H_{\rm cl}-E_F)~,\\
C &=& \frac{g}{(2\pi\hbar)^2}\frac{\hbar^2}{24m}(\nabla_\bR V)^2\int d^2p~ e^{i\bp\cdot\bs/\hbar}\delta''(H_{\rm cl}-E_F)~,\\
D &=& \frac{g}{(2\pi\hbar)^2}\frac{\hbar^2}{24 m^2}\int d^2p~ e^{i\bp\cdot\bs/\hbar}(\bp\cdot\nabla_\bR)^2V \delta''(H_{\rm cl}-E_F)~.
\eea

The analytical evaluation of the above integrals requires the following identities (below, $p_F = \hbar k_F$) , 
\beq\label{delta}
\delta(H_{\rm cl} - E_F) = \frac{m}{p_F}\delta(p-p_F)~,
\eeq
\beq\label{deltaprime}
\frac{d \delta(H_{\rm cl}-E_F)}{d H_{\rm cl}} = \frac{m^2}{p_F p}\frac{d \delta (p-p_F)}{dp}~,
\eeq
and
\beq\label{deltaprimeprime}
\frac{d^2 \delta (H_{\rm cl}-E_F)}{d H_{\rm cl}^2} = \frac{m^3}{p_F}
\left[ \frac{1}{p^2} \frac{d^2 \delta (p-p_F)}{dp^2} - \frac{1}{p^3}\frac{d \delta (p-p_F)}{dp}\right]~.
\eeq
Each of the integrals, $A, B, C, D$,  is explicitly worked out in Appendices~\ref{appA}--\ref{appD}.  Here, we will simply write down our final result for the semi-classical $\hbar$-expansion of the ODM (to ${\cal O}(\hbar^2)$)~\cite{note},
\bea\label{SCfinal}
\bar{\rho}^{\rm sc}_1(\bR;\bs) &=& g\left\{\frac{k_F^2}{2\pi}\frac{J_1(z)}{z} - \frac{1}{48\pi} z J_1(z)
\frac{\nabla^2_\bR k_F^2}{k_F^2} + \frac{1}{96\pi}\frac{z^2J_0(z)}{k_F^2}
\left[\nabla_\bR\left(\nabla_\bR k_F^2\cdot\frac{\bs}{s}\right)\cdot\frac{\bs}{s}\right]\right.\nonumber\\
&+&\left. \frac{1}{192\pi} z^2 J_2(z)\frac{(\nabla_\bR k_F^2)^2}{k_F^4}\right\}~,
\eea
where $z=k_F|\bs|$ and $k_F(\bR) = \sqrt{2m(E_F-V(\bR)/\hbar^2}$ is the local Fermi wave vector.
The first term in $\bar{\rho}^{\rm sc}_1(\bR;\bs)$ agrees exactly with Eq.~\eqref{2DODM},
and highlights that the lowest order contribution to Eq.~\eqref{SCfinal}
corresponds to the LDA, or equivalently, the Thomas-Fermi approximation.  As promised, the GV ODM is manifestly Hermitian, and takes a very different form to the Kirzhnits ODM recently derived
by Putaja {\em et al}.~\cite{Putaja,note3}.

\subsection{The spatial and kinetic energy densities}

The semiclassical spatial density is immediately obtained by taking the diagonal element of Eq.~\eqref{SCfinal}.  However, all terms but the first vanish in taking the
$\bs \to 0$ limit (i.e. $\bR \to \br$), and one obtains,
\beq
\rho^{\rm sc}(\bR) = g \frac{[k_F(\bR)]^2}{4\pi}~,
\eeq
which is just the LDA applied to the uniform gas.   
This result is special to 2D, since in 1D and 3D, non-vanishing gradient corrections are present~\cite{Putaja,kirz,sala,koivisto}.

The KE density may be found by inserting Eq.~\eqref{SCfinal} into Eq.~\eqref{SCKE}, but
upon taking the $\bs \to 0$ limit, all but the LDA term will vanish, leaving
\beq\label{SCKE2}
T_0[\rho] =  \frac{\pi}{g}\frac{\hbar^2}{m}\int d^2R~\rho(\bR)^2 ~.
\eeq
Again, the GV expansion of the ODM has not changed the fact that there are no gradient corrections to the noninteracting KE functional for an inhomogeneous
2D Fermi gas; that is, the KE functional is the Thomas-Fermi functional for a 2D noninteracting Fermi gas which is again unique to 2D systems~\cite{Putaja,kirz,sala,koivisto}.  Gradient corrections to the 2D KE functional
can be motivated within the so-called average density approximation, but this requires the KE functional to be inherently {\em nonlocal}~\cite{vanzylADA}.

\subsection{Consistency criterion of the Euler equation and Idempotency}
Coming first to the consistency criterion established by Gross and Proetto~\cite{gross}, it has already been shown in Ref.~\cite{Putaja} that the Thomas-Fermi KE density functional satisfies  the Euler equation that minimizes the total energy functional, Eq.~\eqref{functional}, 
\beq
\frac{\delta T_0[\rho]}{\delta \rho} + v_{\rm eff}(\br) = E_F~.
\eeq
Owing to the fact that in the GV formulation, only the Thomas-Fermi term survives, consistency is guaranteed.

The idempotency of the GV semiclassical ODM  has already been proved in  arbitrary dimensions in Ref.~\cite{BR}.  It follows that our 2D ODM,  Eq.~\eqref{SCfinal}, is also idempotent, which is in fact a strong
constraint to be placed on any approximate density matrix.  


\section{Application: 2D spin-polarized dipolar Fermi gas}
In this section, we will use our semiclassical expansion for the ODM, Eq.~\eqref{SCfinal}, to go beyond
the LDA for the total dipole-dipole interaction energy functional of a spin-polarized
(all moments aligned parallel with the
$z$-axis),
inhomogeneous 2D dipolar Fermi gas.
We restrict ourselves to the HFA, where the total dipolar interaction energy is given by
\beq\label{Eint}
E_{\rm int} = \frac{1}{2}\int d^2r \int d^2r'~ [\rho_1(\br;\br)\rho_1(\br';\br') - \rho_1(\br;\br')\rho_1(\br';\br)]
V_{dd}(\br-\br')~,
\eeq
and
\beq\label{Vdd}
V_{dd}(\br-\br') = \frac{\mu_0 d^2}{4\pi}\frac{1}{|\br-\br'|^3}~,
\eeq
is the interaction potential between two magnetic dipoles restricted to locations $\br$ and
$\br'$ in the 2D $xy$-plane,  and $d$ is the magnetic moment of an atom.  The individual terms
in Eq.~\eqref{Eint} are the direct and exchange energies, and while they are separately divergent for a $1/r^3$ potential in 2D, their sum is finite owing to the Pauli
exclusion principle~\cite{vanzyl_pisarski}.  As discussed at length in Ref.~\cite{vanzyl_pisarski},
it is convenient to work with a {\em regularized} dipolar interaction, which leads to the following
expression for the total interaction energy within the HFA (details of this calculation have already
been presented in Ref.~\cite{vanzyl_pisarski}),
\bea\label{Eint2}
E_{\rm int} &=& \frac{\mu_0 d^2}{8\pi} \int d^2s~\frac{1}{s^3}[f(0)-f(\bs)] -\frac{\mu_0 d^2}{4} \int \frac{d^2q}{(2\pi)^2}~q |\tilde{\rho}(\bq)|^2 
\nonumber \\
&\equiv& E^{(1)}_{dd} + E^{(2)}_{dd}~.
\eea
where $\tilde{\rho}(\bq)$ is the 2D Fourier transform of the density, and
\beq\label{fs}
f(0)-f(\bs) = \int d^2R\{[\bar{\rho}_1(\bR;0)]^2 - [\bar{\rho}_1(\bR;\bs)]^2\}~.
\eeq
 It is important to emphasize here that $E^{(1)}_{dd}$ and $E^{(2)}_{dd}$ are {\em not} to be
interpreted as the direct and exchange energies, respectively.
Note that $E^{(2)}_{dd}$ is the nonlocal contribution to the HF energy, and as written is exact.  On the
other hand, in order to get an explicit expression for $E^{(1)}_{dd}$ in terms of the density, we
need to invoke some level of approximation to $f(0)-f(\bs)$.

To this end, we define
the radial distribution function for the inhomogeneous system as
\beq\label{gs}
g(\bR;\bs) = 1 - \frac{[\bar{\rho}_1(\bR;\bs)]^2}{[\bar{\rho}_1(\bR;0)]^2}~,
\eeq
so that we may write
\beq\label{E1dd}
E^{(1)}_{dd}= \frac{\mu_0 d^2}{8\pi}\int d^2R \int d^2s \frac{1}{s^3}[\rho(\bR)]^2 g(\bR;\bs)~.
\eeq
Now, taking only the leading order term from our semiclassical expansion of the ODM, Eq.~\eqref{SCfinal}
($g=1$), we immediately obtain
\bea\label{fsLDA}
E^{(1),{\rm LDA}}_{dd} &=& \frac{\mu_0 d^2}{4} \int d^2R [\rho(\bR)]^2k_F(\bR)
\int_0^\infty dz \frac{1}{z^2}\left[1-\left(\frac{2 J_1(z)}{z}\right)^2\right]\nonumber \\
&=& \mu_0 d^2\frac{\sqrt{\pi}}{2}\int d^2R [\rho(\bR)]^{5/2}\frac{128}{45\pi}\nonumber \\
&=& \mu_0 d^2\frac{64}{45\sqrt{\pi}}\int d^2R~[\rho(\bR)]^{5/2}~,
\eea
which is in perfect agreement with Eq.~(27) in Ref.~\cite{vanzyl_pisarski}.

We may now go beyond the LDA for $E^{(1)}_{dd}$ by taking in turn all of the $\hbar^2$ corrections
to the ODM in Eq.~\eqref{SCfinal}.  To begin, we note that to $ {\cal O}(\hbar^2)$
\bea\label{ODMsq}
| \bar{\rho}_1^{\rm sc}(\bR;\bs)|^2 &=& \frac{k_F^4}{4\pi^2}\left(\frac{J_1(z)}{z}\right)^2 +
\frac{z J_1(z)J_2(z)}{192 \pi^2} \frac{(\nabla_\bR k_F^2)^2}{k_F^2} - \frac{(J_1(z))^2}{48 \pi^2}
(\nabla_\bR^2 k_F^2)\nonumber \\
&+&\frac{zJ_1(z)J_0(z)}{96 \pi^2}\left[\nabla_\bR\left(\nabla_\bR k_F^2\cdot\frac{\bs}{s}\right)
\cdot\frac{\bs}{s}\right]~,
\eea
from which we obtain
\bea\label{fsf0}
g^{\rm sc}(\bR;\bs) &=& 1-\frac{[\bar{\rho}_1^{\rm sc}(\bR;\bs)]^2}{[\rho(\bR)]^2} \nonumber \\
&=& \left[1-\left(\frac{2 J_1(z)}{z}\right)^2\right] -\frac{zJ_1(z)J_2(z)}{48\pi} \frac{(\nabla_\bR\rho)^2}{\rho^3}+\frac{1}{48\pi^2}\frac{\nabla^2_\bR k^2_F}{\rho^2}[J_1(z)]^2\nonumber \\
&-& \frac{z J_1(z)J_0(z)}{96 \pi^2}\frac{1}{\rho^2}\left[\nabla_\bR\left(\nabla_\bR k_F^2\cdot\frac{\bs}{s}\right)
\cdot\frac{\bs}{s}\right]~.
\eea
We may now write
\bea\label{EintSO}
E^{(1),+}_{dd} &=& \frac{\mu_0 d^2}{8 \pi}\int d^2R\int d^2s~\frac{1}{s^3}[\rho(\bR)]^2 g^{\rm sc}(\bR;\bs) \nonumber \\
&=& \mu_0 d^2\frac{\sqrt{\pi}}{2}\int d^2R [\rho(\bR)]^{5/2}\int_0^{\infty} dz~\frac{1}{z^2}\left\{\left[1-\left(\frac{2 J_1(z)}{z}\right)^2\right]-\frac{zJ_1(z)J_2(z)}{48\pi} \frac{(\nabla_\bR\rho)^2}{\rho^3}\right.\nonumber \\
&+&\left. \frac{1}{48\pi^2}\frac{\nabla^2_\bR k^2_F}{\rho^2}[J_1(z)]^2\right\}
-\frac{\mu_0 d^2}{8\pi} \int d^2R~\int d^2 s~\frac{1}{s^3} \frac{z J_z(z)J_0(z)}{96 \pi^2}\left[\nabla_\bR\left(\nabla_\bR k_F^2\cdot\frac{\bs}{s}\right)\right]~.
\eea
The first term in square braces in Eq.~\eqref{EintSO} has already been shown to yield the LDA
to $E^{(1)}_{dd}$, viz., Eq.~\eqref{fsLDA}.  Let us consider now the second term, defined by
\bea\label{I2}
I_2 &=& -\mu_0 d^2\frac{\sqrt{\pi}}{2} \int d^2R~[\rho(\bR)]^{5/2} \int_0^{\infty} dz~\frac{1}{z} \frac{J_1(z)J_2(z)}{48\pi} \frac{(\nabla_\bR\rho)^2}{\rho^3}\nonumber\\
&=&  -\mu_0 d^2\frac{\sqrt{\pi}}{2} \int d^2R~[\rho(\bR)]^{5/2} \frac{1}{72\pi^2} \frac{(\nabla_\bR\rho)^2}{\rho^3}~,
\eea
where Mathematica$^\copyright$ has been used to evaluate the $z$-integral.
The third integral to be evaluated is
\bea\label{I3}
I_3 &=&\mu_0 d^2\frac{\sqrt{\pi}}{2} \int d^2R~[\rho(\bR)]^{5/2} \frac{1}{48\pi^2}\frac{\nabla^2_\bR k^2_F}{\rho^2} \int_0^{\infty} dz~\frac{1}{z^2}
[J_1(z)]^2\nonumber \\
&=&\mu_0 d^2\frac{\sqrt{\pi}}{2} \int d^2R~[\rho(\bR)]^{5/2} \frac{1}{36\pi^3} \frac{\nabla^2_\bR k^2_F}{\rho^2}\nonumber \\
&=&\mu_0 d^2\frac{\sqrt{\pi}}{2} \int d^2R~[\rho(\bR)]^{5/2} \frac{1}{9 \pi^2}\frac{\nabla_\bR^2\rho}{\rho^2}\nonumber \\
&=&-\mu_0 d^2\frac{\sqrt{\pi}}{2} \int d^2R~[\rho(\bR)]^{5/2} \frac{1}{18 \pi^2}\frac{(\nabla_\bR\rho)^2}{\rho^3}~.
\eea
In going to the last line in Eq.~\eqref{I3}, we have assumed that the spatial density vanishes at infinity, viz., $\rho(\bR \to \infty) = 0$.

The last integral in Eq.~\eqref{EintSO} is more involved, so we leave the details to Appendix~\ref{appE}.  The result of this calculation is given by
\bea\label{I4}
I_4 &=& -\frac{\mu_0 d^2}{8\pi} \int d^2R~\int d^2 s~\frac{1}{s^3} \frac{z J_z(z)J_0(z)}{96 \pi^2}\left[\nabla_\bR\left(\nabla_\bR k_F^2\cdot\frac{\bs}{s}\right)
\cdot\frac{\bs}{s}\right]\nonumber \\
&=& \mu_0 d^2 \frac{\sqrt{\pi}}{2}\int d^2R~[\rho(\bR)]^{5/2}\left(\frac{1}{48 \pi^2}\frac{(\nabla_\bR\rho)^2}{\rho^3}\right)~.
\eea
Summing all of the contributions finally leads to
\bea\label{ELDAplus}
E^{(1),+}_{dd} &=& \mu_0 d^2\frac{\sqrt{\pi}}{2}\int d^2R~[\rho(\bR)]^{5/2}\left[\frac{128}{45\pi} -\frac{1}{72\pi^2}
\frac{(\nabla_\bR\rho)^2}{\rho^3}-\frac{1}{18\pi^2}\frac{(\nabla_\bR\rho)^2}{\rho^3} + \frac{1}{48\pi^2}\frac{(\nabla_\bR\rho)^2}{\rho^3}\right]\nonumber\\
&=& \mu_0 d^2\frac{\sqrt{\pi}}{2}\int d^2R~[\rho(\bR)]^{5/2}\left[\frac{128}{45\pi} -\frac{7}{144 \pi^2}\frac{(\nabla_\bR\rho)^2}{\rho^3} \right]~.
\eea
It is a little surprising that the second order correction to $E^{(1),+}_{dd}$ tends to {\em lower} the energy, but we need to remember that $E^{(1)}_{dd}$ is part
of the {\em total} dipole-dipole energy, in which case, the expected sign of the correction may be not fit in with our intuition.  Moreover, the coefficient in front of the correction term
is $\sim - 0.005$, which suggests that it is a small contribution relative to the first LDA term.

One way to quantify the improvement of the gradient correction is to consider the relative percentage error
(RPE) between the approximate value of $E^{(1)}_{dd}$, and the exact value, viz.,
\beq\label{RPE}
{\rm RPE} \equiv \frac{|E^{(1), \rm{approx}}_{dd} - E^{(1),{\rm ex}}_{dd}|}{E^{(1),{\rm ex}}_{dd}}\times 100~,
\eeq
where in Eq.~\eqref{RPE}, $E^{(1),{\rm approx}}_{dd}$ is either $E^{(1),{\rm LDA}}_{dd}$ or $E^{(1),+}_{dd}$

It  has already been shown in Ref.~\cite{vanzyl_pisarski} that the {\em exact}
expression, $E^{(1),{\rm ex}}_{dd}$, for a spin-polarized, harmonically confined 2D Fermi gas  is given by (here, energies are scaled by $\mu_0 d^2/a^3_{\rm ho}$ and lengths by
$a_{\rm ho} = \sqrt{\hbar/m\omega_0}$, where $\omega_0$ is the trap frequency),
\bea\label{E1ex}
E^{(1),{\rm ex}}_{dd} &=& \frac{1}{4 \pi}\frac{1}{\sqrt{2}}\sum_{n=0}^M \frac{(n+1)\Gamma(n+3/2)}{\Gamma(n+1)}
\left\{\frac{4}{3} n~ _3 F_2\left(-\frac{3}{2},\frac{1}{2},-n;2,-\frac{1}{2}-n;1\right)\right. \nonumber \\ 
&+& \left. _3 F_2\left(-\frac{1}{2},\frac{1}{2},-n;2,-\frac{1}{2}-n;1\right)   \right\}
\eea
where we have assumed $M+1$ closed shells, and the particle number is given by $N = \frac{1}{2}(M+1)(M+2)$. Note that here,  by exact, we mean that the exact ODM for the harmonic oscillator, Eq.~\eqref{ODMex} below, has been used to evaluate Eq.~\eqref{Eint}.
 We also have the exact ODM, which reads~\cite{vanzyl2003,vanzyl_brack}
\beq\label{ODMex}
\bar{\rho}^{\rm ex}_1(\bR;\bs) = \frac{1}{\pi}\sum_{n=0}^{M} (-1)^n L_n(2 R^2) L^{1}_{M-n}(s^2/2) e^{-(R^2 + s^2/4)}
\eeq
from which the exact density is given by taking $\bs=0$ in Eq.~\eqref{ODMex}, viz.,
\beq\label{denex}
\bar{\rho}^{\rm ex}_1(\bR;0) =\rho_{\rm ex}(\bR) = \frac{1}{\pi}\sum_{n=0}^M (-1)^n (M-n+1) L_n(2 R^2) e^{-R^2}~,
\eeq
where $_3 F_2[a,b,c;d,e;z]$ is a generalized Hypergeometric function, and $L^\alpha_n(z)$ is a Laguerre polynomial~\cite{GR}.  Inserting Eq.~\eqref{denex} into Eq.~\eqref{fsLDA} gives the second
column in Table~\ref{table1}, while inserting Eq.~\eqref{denex} into Eq.~\eqref{ELDAplus} yields the third column.  We have focused our attention to $N\sim 50-500$ particles since it is in this regime
where we expect the most significant deviations from the exact results.
It is clear from Table~\ref{table1} that the negative correction serves to bring $E^{(1),+}_{dd}$ and $E^{(1),{\rm ex}}_{dd}$  into much closer agreement.  The RPE for $E^{(1),{\rm LDA}}_{dd}$ and $E^{(1),+}_{dd}$ are displayed in 
the fifth and sixth 
columns of Table.~\ref{table1}, respectively.  We note that for $N=55$, the gradient correction already reduces the RPE by
a factor of three, highlighting that while the correction is small, it significantly improves the agreement with the
exact result given by Eq.~\eqref{E1ex}.
In this sense, the negative sign of the gradient correction in Eq.~\eqref{ELDAplus} is justified {\em a posteriori}.
\begin{table}[ht] 
\centering      
\begin{tabular}{c c c c  c c }  
\hline\hline                        
$N$ & $E^{(1),{\rm LDA}}_{dd}$ & $E^{(1),+}_{dd}$ & $E^{(1),{\rm ex}}_{dd}$ & RPE$^{{\rm LDA}}$ & RPE$^{+}$ \\ [0.5ex] 
\hline                    
55    &  54.5725   &  54.4547  & 54.4003 & 0.3 & 0.1 \\
105  &  168.937   &  168.739  & 168.654 &0.2& 0.05   \\
231    &  670.718   &   670.350    & 670.199  & 0.08& 0.02 \\
496   &  2553.50   &  2552.83  & 2552.48  & 0.04 & 0.01 \\
[1ex]       
\hline     
\end{tabular} 
\caption{A comparision of the LDA (second column), gradient corrected (third column) and exact (fourth column) expressions for $E^{(1)}_{dd}$.  The last two columns correspond to the RPE defined in Eq.~\eqref{RPE}.  Energies are in units of $\mu_0 d^2/a^3_{\rm ho}$,
as discussed in the text.} 
\label{table1}  
\end{table} 

\begin{figure}[ht]
\centering \scalebox{0.5}
{\includegraphics{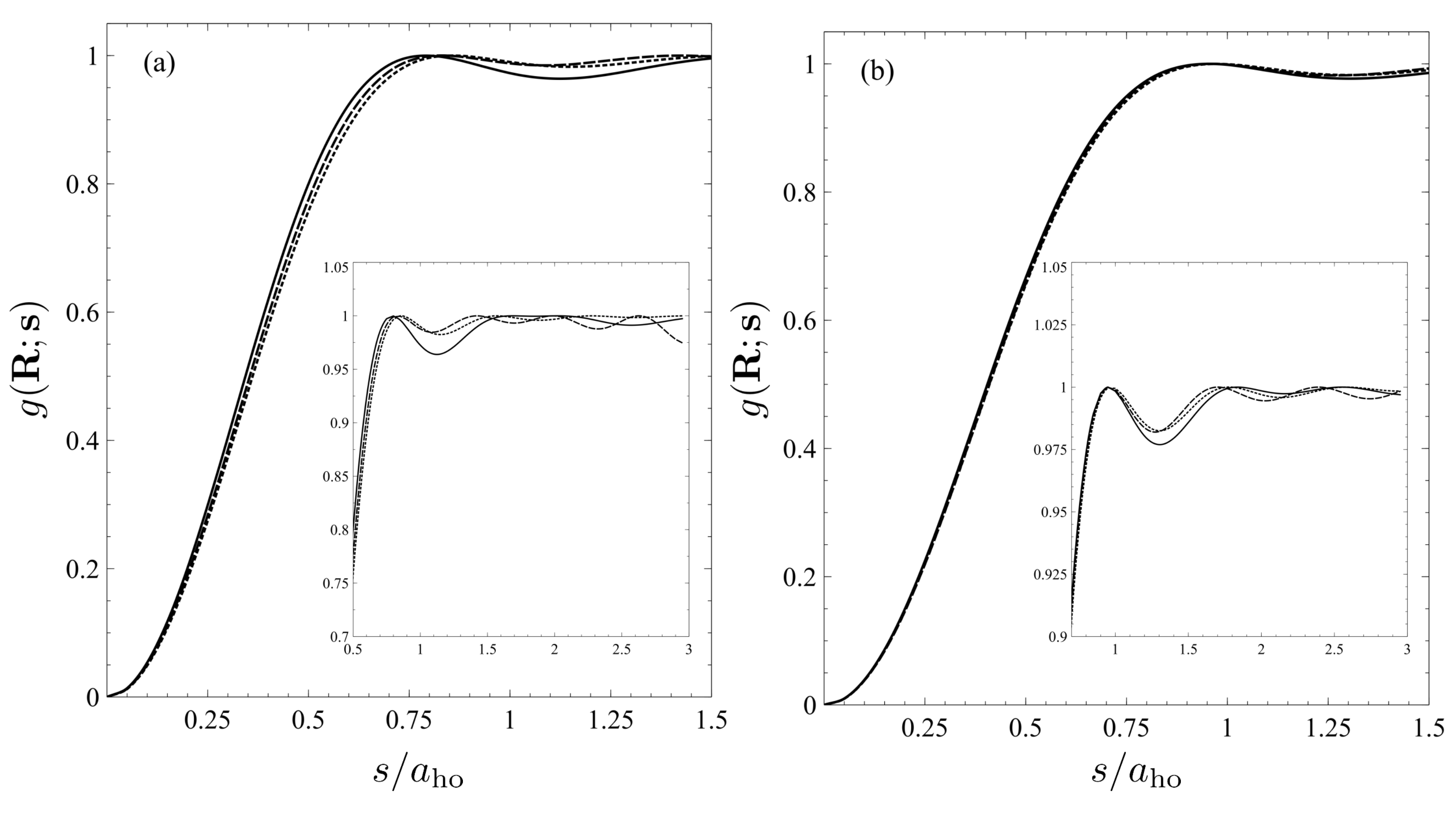}}
\caption{Solid, dashed and dotted curves correspond to the radial distribution functions, $g^{\rm ex}(R;s)$, $g^{\rm sc}(R;s)$ and $g^{\rm LDA}(R;s)$, respectively.  Panel (a) is evaluated for $R=0$, and panel
(b) is for $R=\frac{1}{2}R_{\rm TF}$, where the Thomas-Fermi radius is given by $R_{\rm TF}/a_{\rm ho}=\sqrt{2\sqrt{2 N} }$ with $N=55$ particles. 
As defined by Eq.~\eqref{gs}, $g(\bR;\bs)$  is dimensionless. The insets to both panels depict a zoomed-in, extended range for the radial
distribution functions.  The axes of the insets are as in the main figure.}
\label{fig1}
\end{figure}

The energies presented in Table~\ref{table1} provide a global comparision, in the sense that they are integrated quantities.  Another useful test to understand why the gradient corrections to the ODM, Eq.~\eqref{SCfinal},
provide such an improvement  to
the HF energy, is to 
consider a {\em pointwise} comparision (i.e., a local comparision) of the radial distribution functions described by the exact, $g^{\rm ex}(R;s)$, gradient corrected, $g^{\rm sc}(R;s)$,  and LDA, $g^{\rm LDA}(R;s)$.
In Fig.~\ref{fig1} we present two panels, which display the exact (solid curve), gradient corrected (dashed curve) and the LDA (dotted curve) radial distribution functions.  Panel (a) is evaluated at $R=0$, where the
largest discrepancy between the distributions is present.  It is clear that the inclusion of gradient corrections brings $g^{\rm sc}(R;s)$ into closer agreement with $g^{\rm ex}(R;s)$ for
$s/a_{\rm ho}<1$. In Panel (b), we evaluate the radial distributions at $R/R_{\rm TF}=\frac{1}{2}$, where we observe that all three distributions are in very good agreement for $s/a_{\rm ho}<1$.
 The insets to both panels show a zoomed in, extended range for the distribution functions.   It is evident that for $s/a_{\rm ho}>1$, both $g^{\rm sc}(R;s)$ and $g^{\rm LDA}(R;s)$ over-estimate and under-estimate the exact distribution in
an oscillatory fashion.
Since $E^{(1)}_{dd}$ involves the integration over $\bR$ and $\bs$, the oscillatory under-estimation and over-estimation of the distributions tends to average out, with the
net result that both $E^{(1),+}_{dd}$  and $E^{(1),{\rm LDA}}_{dd}$ remain close to the exact value, $E^{(1),{\rm ex}}_{dd}$.

\section{Summary}
We have applied the semiclassical $\hbar$-expansion of Grammaticos and Voros to construct a manifestly Hermitian,
idempotent, one-body density matrix for a two-dimensional Fermi gas to second-order
in $\hbar$.  While our density matrix also satisfies the consistency criterion of the Euler
equation, it does not remedy the fact that in two-dimensions, the noninteracting 
kinetic energy functional
has vanishing gradient corrections to all orders in $\hbar$.  

As an interesting application, we have provided a detailed calculation for the second-order correction
to the Hartree-Fock energy of a spin-polarized, two-dimensional dipolar Fermi gas.  
We find a small, but finite, negative gradient correction 
to the local-density approximation.  To test the quality of the correction,
we have performed numerical comparisons with the known exact results for a harmonically confined,
spin-polarized, two-dimensional Fermi gas.  We find that including the gradient
correction yields superlative agreement with the exact dipole-dipole interaction energy, at least for the case of harmonic confinement.

There are several areas of research where the results of this paper may be useful.  One could
use our beyond local-density approximation for the total dipole-dipole interaction energy in a
density-functional theory application for the equilibrium, collective properties, and
density instabilities, of a spin-polarized two-dimensional dipolar Fermi gas.  We also see potential
applications of our one-body density matrix for developing gradient corrected 
interaction energy density functionals in 
inhomogeneous, 
two-dimensional degenerate
electronic systems, which could be used in, e.g., density-functional theory studies of
two-dimensional quantum dots.   Finally, it would be of interest to consider our
semiclassical expansion to higher order in $\hbar$, so that we may ascertain if such corrections remain finite, and
if so, to determine whether the semiclassical expansion is convergent or asymptotic.

\acknowledgements
This work was supported by grants from the Natural Sciences and Engineering Research Council of Canada (NSERC).


\appendix
\section{}\label{appA}

In the following we shall evaluate Eq.~(22).
\bea
A &=& \frac{g}{(2\pi\hbar)^2}\int_0^{p_F} dp~p\int_0^{2\pi} d\phi~e^{ips\cos(\phi)/\hbar}\nonumber\\
&=& \frac{g}{2\pi s^2}\int_0^z du~uJ_0(u)~,
\eea
where we have put $u=ps/\hbar$ and $z=k_Fs$.  The resulting integral can be performed using Mathematica$^\copyright$ and, gives
\beq\label{A}
A = g\frac{k_F^2}{2\pi}\frac{J_1(z)}{z}
\eeq
\section{}\label{appB}
The evaluation of Eq.~(23) proceeds as follows.
\bea
B &=& \frac{g}{(2\pi\hbar)^2}\frac{\hbar^2}{8m}\nabla^2_\bR V\int_0^\infty dp~p\delta'(H_{\rm cl}-E_F)\int_0^{2\pi}d\phi~e^{ips\cos(\phi)/\hbar}\nonumber\\
&=& g\frac{2\pi}{(2\pi\hbar)^2}\frac{\hbar^2}{8m}\nabla^2_\bR V\int_0^\infty dp~p\delta'(H_{\rm cl}-E_F)J_0\left(\frac{ps}{\hbar}\right)~.
\eea
Using Eq.~\eqref{deltaprime}, we can write
\bea
B&=& g \frac{m}{16\pi p_F}\nabla^2_\bR V \int_0^\infty dp~\frac{d \delta(p-p_F)}{dp}J_0\left(\frac{ps}{\hbar}\right)\nonumber \\
&=& g\frac{m}{16\pi\hbar^2 k_F^2}\nabla^2_\bR V(\bR)  z J_1(z)~,
\eea
where we have used $\frac{d J_0(x)}{dx}=-J_1(x)$.  Finally, we may use $\nabla^2_\bR V(\bR) = -\frac{\hbar^2}{2m}\nabla_\bR^2 k_F^2$ to write
\bea\label{B}
B = -g\frac{1}{32 \pi}\frac{\nabla_\bR^2 k_F^2}{k_F^2}~z J_1(z)~.
\eea
\section{}\label{appC}
The evaluation of Eq.~(24) may be performed if we write
\bea
C &=& g \frac{1}{48\pi m}(\nabla^2_\bR V)^2 \int_0^\infty dp~p \delta''(H_{\rm cl}-E_F) J_0\left(\frac{ps}{\hbar}\right)~.
\eea
Upons substituting Eq.~\eqref{deltaprimeprime}, we obtain
\bea
C &=& g \frac{m^2}{48 \pi p_F} (\nabla^2_\bR V)^2\int_0^\infty dp~\left[ \frac{1}{p}\frac{d^2\delta(p-p_F)}{dp^2} - \frac{1}{p^2}\frac{d\delta(p-p_F}{dp}\right] J_0\left(\frac{ps}{\hbar}\right)\nonumber \\
&=&  g \frac{m^2}{48 \pi p_F} (\nabla^2_\bR V)^2\left\{ \frac{d^2}{dp^2}\left[\frac{1}{p} J_0\left(\frac{ps}{\hbar}\right)\right] + \frac{d}{dp}\left[\frac{1}{p^2} J_0\left(\frac{ps}{\hbar}\right)\right]\right\}_{p=p_F}~.
\eea
Recalling that $z=p_Fs/\hbar=k_F s$, we can write
\bea\label{tempC}
C &=& g \frac{m^2}{48 \pi p_F} (\nabla^2_\bR V)^2\frac{s^3}{\hbar^3}\left\{\frac{d^2}{du^2}\left[\frac{J_0(u)}{u}\right] + \frac{d}{du}\left[\frac{J_0(u)}{u^2}\right]\right\}_{u=z}\nonumber \\
&=& g \frac{m^2}{48 \pi \hbar^4 k_F^4}(\nabla_\bR V)^2 z^3 
\left\{\frac{d^2}{dz^2}\left[\frac{J_0(z)}{z}\right] + \frac{d}{dz}\left[\frac{J_0(z)}{z^2}\right]\right\}~.
\eea
Performing the derivatives with respect to $z$ gives after simplification (we have used Mathematica$^\copyright$),
\beq\label{derident}
\frac{d^2}{dz^2}\left[\frac{J_0(u)}{z}\right] + \frac{d}{dz}\left[\frac{J_0(z)}{z^2}\right] = \frac{J_2(z)}{z}~.
\eeq
Substituting Eq.~\eqref{derident} into Eq.~\eqref{tempC}, we get
\bea\label{C}
C &=& g \frac{m^2}{48 \pi \hbar^4 k_F^4}(\nabla_\bR V)^2 z^2 J_2(z) \nonumber \\
&=& g \frac{1}{192 \pi} z^2 J_2(z) \frac{(\nabla_\bR k_F^2)^2}{k_F^4}~,
\eea
where in going to the last line in Eq.~\eqref{C}, we have made use of $\nabla_\bR V = -\frac{\hbar^2}{2m}(\nabla_\bR k_F^2)$.
\section{}\label{appD}
Equation~(25) is the most difficult to evaluate, and requires some care.  Let us first rewrite Eq.~(25) in the following form
\bea
D &=& \frac{g}{(2\pi\hbar)^2} \frac{\hbar^2}{24m^2}\sum_{i=1}^2\sum_{j=1}^2 \left[\frac{\partial^2 V}{\partial X_i\partial X_j}\right]\int d^2p~ p_i p_j e^{i\bp\cdot\bs/\hbar}
\delta''(H_{\rm cl}-E_F)~.
\eea
Next, we make use of the identity
\beq
p_i p_j  e^{i\bp\cdot\bs/\hbar} = -\hbar^2 \frac{\partial^2  e^{i\bp\cdot\bs/\hbar}}{\partial s_i \partial s_j}~,
\eeq
which allows us to write
\bea
D &=& -g\frac{\hbar^2}{96 \pi^2 m^2}\sum_{i=1}^2\sum_{j=1}^2 \left[\frac{\partial^2 V}{\partial X_i\partial X_j}\frac{\partial^2}{\partial s_i\partial s_j}  \right]
\int d^2p~e^{i\bp\cdot\bs/\hbar}
\delta''(H_{\rm cl}-E_F)\nonumber\\
&=& -g\frac{\hbar^2}{48 \pi  m^2}\sum_{i=1}^2\sum_{j=1}^2 \left[\frac{\partial^2 V}{\partial X_i\partial X_j}\frac{\partial^2}{\partial s_i\partial s_j}  \right] \int_0^\infty dp~p~
\delta''(H_{\rm cl}-E_F) J_0\left(\frac{ps}{\hbar}\right)\nonumber \\
&=& -g\frac{\hbar^2 m}{48 \pi p_F}\sum_{i=1}^2\sum_{j=1}^2 \left[\frac{\partial^2 V}{\partial X_i\partial X_j}\frac{\partial^2}{\partial s_i\partial s_j}  \right] \int_0^\infty dp~
\left[ \frac{1}{p}\frac{d^2\delta(p-p_F)}{dp^2} - \frac{1}{p^2}\frac{d\delta(p-p_F}{dp}\right] J_0\left(\frac{ps}{\hbar}\right)~,
\eea
and proceeding as we did for the evaluation of $C$, we arrive at
\bea\label{tempD}
D &=& -g \frac{m}{48 \pi \hbar^2 k_F^4} \sum_{i=1}^2\sum_{j=1}^2 \left[\frac{\partial^2 V}{\partial X_i\partial X_j}\frac{\partial^2}{\partial s_i\partial s_j}  \right] z^2 J_2(z)~.
\eea
Let us now define
\beq
U \equiv \frac{\partial^2}{\partial s_i \partial s_j}(z^2 J_2(z))~.
\eeq
Once again, using $z=p_Fs/\hbar$, we obtain
\bea
U &=& \frac{p_F}{\hbar}\frac{\partial}{\partial s_i}\left[\frac{s_j}{s}\frac{\partial(z^2 J_2(z))}{\partial z}\right]\nonumber \\
&=& \frac{p_F}{\hbar}\left[ \frac{\delta_{ij}}{s}\frac{\partial(z^2J_2)}{\partial z} - \frac{s_i s_i}{s^3}\frac{\partial (z^2 J_2)}{\partial z} +
\frac{s_j}{s}\frac{\partial^2 (z^2 J_2)}{\partial z^2}\frac{\partial z}{\partial s_i}\right]\nonumber \\
&=& \frac{p_F}{\hbar}\left[ \frac{\delta_{ij}}{s}\frac{\partial(z^2J_2)}{\partial z} - \frac{s_i s_i}{s^3}\frac{\partial (z^2 J_2)}{\partial z} +
\frac{p_F}{\hbar}\frac{s_i s_j}{s^2}\frac{\partial^2 (z^2 J_2)}{\partial z^2}\right]~.
\eea
Finally, making use of the readily derived identity
\beq
\frac{\partial^2 (z^2 J_2)}{\partial z^2} = z J_1(z) + z^2 J_0(z)~,
\eeq
we obtain after some straightforward simplification
\bea\label{U}
U = k_F^2\left[ \delta_{ij} z J_1(z) + \frac{s_i s_j}{s^2} z^2 J_0(z)\right]~.
\eea
Using our expression for $U$, Eq.~\eqref{U}, in Eq.~\eqref{tempD}, we finally arrive at
\bea\label{D}
D &=& -g  \frac{m}{48 \pi \hbar^2 k_F^2}\sum_{i=1}^2\sum_{j=1}^2 \frac{\partial^2 V}{\partial X_i\partial X_j} \left[ \delta_{ij} z J_1(z) + \frac{s_i s_j}{s^2} z^2 J_0(z)\right] \nonumber \\
&=& -g  \frac{m}{48 \pi \hbar^2 k_F^2}\sum_{i=1}^2\sum_{j=1}^2 \frac{\partial^2 V}{\partial X_i\partial X_j} \delta_{ij} zJ_1(z) -g  \frac{m}{48 \pi \hbar^2 k_F^2}\sum_{i=1}^2\sum_{j=1}^2 \frac{\partial^2 V}{\partial X_i\partial X_j}
\left[\frac{s_i s_j}{s^2}z^2 J_0(z)\right]\nonumber \\
&=& -g \frac{m}{48 \pi \hbar^2 k_F^2}(\nabla^2_\bR V) z J_1(z)  -g \frac{m}{48 \pi \hbar^2 k_F^2}
\left[\nabla_\bR \left( \nabla_\bR V \cdot \frac{\bs}{s}\right)\cdot \frac{\bs}{s}\right] z^2 J_0(z)\nonumber \\
&=& g \frac{1}{96 \pi} \frac{\nabla^2_\bR k_F^2}{k_F^2} z J_1(z) +  g  \frac{1}{96\pi}\frac{z^2J_0(z)}{k_F^2}
\left[\nabla_\bR\left(\nabla_\bR k_F^2\cdot\frac{\bs}{s}\right)\cdot\frac{\bs}{s}\right]~.
\eea
Adding the terms $A+B+C+D$ gives Eq.~\eqref{SCfinal}.
\section{}\label{appE}
We wish to evaluate the following integral,
\bea\label{appI4}
I &=&  \int d^2R~\int d^2 s~\frac{1}{s^3} \frac{z J_z(z)J_0(z)}{96 \pi^2}\left[\nabla_\bR\left(\nabla_\bR k_F^2\cdot\frac{\bs}{s}\right)
\cdot\frac{\bs}{s}\right]~.
\eea
Let us start by presenting what will prove to be a useful expression,
\beq\label{nablaR}
\nabla_\bR = \frac{z}{2 k_F^2}\left(\nabla_\bR k_F^2)\right)\frac{d}{dz}~,
\eeq
which along the $i$-th direction reads
\beq\label{ithnabla}
\frac{\partial}{\partial X_i} = \frac{z}{2 k_F^2}\left(\frac{\partial k_F^2}{\partial X_i}\right) \frac{d}{dz}~.
\eeq
Now, we write Eq.~\eqref{appI4} as
\bea\label{appI42}
I&=& \frac{1}{96 \pi^2}\sum_{i=1}^2\sum_{j=1}^2\int d^2s~\frac{s_i s_j}{s^5} \int d^2R~\left[\frac{\partial^2 k_F^2}{\partial X_i \partial X_j}(z J_1(z) J_0(z))\right]\nonumber \\
&=& -\frac{1}{96 \pi^2}\sum_{i=1}^2\sum_{j=1}^2\int d^2s~\frac{s_i s_j}{s^5} \int d^2R~\left[\frac{\partial k_F^2}{\partial X_j}\frac{\partial(z J_1(z) J_0(z))}{\partial X_i}\right]~.
\eea
Utilizing Eq.~\eqref{ithnabla} in Eq.~\eqref{appI42}, one obtains
\bea\label{appI43}
I &=& -\frac{1}{192 \pi^2}\sum_{i=1}^2\sum_{j=1}^2\int d^2R \frac{1}{k_F^2} \left[\frac{\partial k_F^2}{\partial X_j}\frac{\partial k_F^2}{\partial X_i}\right]
\int_0^{\infty} ds~ \frac{z}{s^4} \frac{d}{dz}[(zJ_1(z) J_0(z))]\int_0^{2\pi} d\phi~ s_i s_j\nonumber \\
&=& -\frac{1}{192\pi}\int d^2R~ \frac{(\nabla_\bR k_F^2)^2}{k_F} \int_0^\infty dz~\frac{1}{z} \frac{d}{dz}[(zJ_1(z) J_0(z))]~,
\eea
where we have used
\beq
\int_0^{2\pi} d\phi~s_i s_j = \pi s^2 \delta_{ij}~.
\eeq
The $z$-integral in Eq.~\eqref{appI43} can be computed using Mathematica$^\copyright$, and evaluates to $2/\pi$, whence we obtain
\bea\label{appI44}
I &=& -\frac{1}{96 \pi^2}\int d^2R~\frac{(\nabla_\bR k_F^2)^2}{k_F}\nonumber \\
&=& -\frac{1}{12\sqrt{\pi}}\int d^2R~[\rho(\bR)]^{5/2}\frac{(\nabla_\bR\rho)^2}{\rho^3}~.
\eea
Upon taking into account the $-\mu_0 d^2/8\pi$ factor in Eq.~\eqref{I4}, we finally arrive at
\bea
I_4 &=& -\frac{\mu_0 d^2}{8\pi}I= \mu_0 d^2 \frac{\sqrt{\pi}}{2}\int d^2R~[\rho(\bR)]^{5/2}\left(\frac{1}{48 \pi^2}\frac{(\nabla_\bR\rho)^2}{\rho^3}\right)~.
\eea


\begin{thebibliography}{99}
\bibitem{note1}
Note that another common convention in the literature is to write $\delta'(E_F - H_{\rm cl})$ and $\delta''(E_F - H_{\rm cl})$, in which case the
sign of the second term in Eq.~\eqref{WT2} becomes positive.  Identical results are, of course, obtained for the ODM regardless of the notational convention.
\bibitem{note}
We have confirmed that by applying the Wigner-Kirkwood (WK) formalism, one obtains an identical 
$\hbar$-expansion for the ODM.  However, in the WK approach, the starting point is the Wigner transform of the 
one-body Bloch density matrix (sometimes called the propagator), which can be related back
to the ODM via an inverse Laplace transform, and yields the ODM in terms of $\bR$ and $\bs$.
See Ref.~\cite{brack_bhaduri}, Ch. 4.
\bibitem{note3}
In the Kirzhnits approach, the ODM is expressed, from the beginning, using $\br$ and $\br'$.  In this sense, the Kirzhnits ODM does not
display the same explicit symmetry as the GV approach, which is formulated in terms of $\bR$ and $\bs$.  We have shown (to be presented in a
future publication) that the Kirzhnits ODM is identical to the GV ODM upon an appropriate symmetrization procedure.
\end{thebibliography}
\end{document}